\documentclass[pdftex,12pt,a4paper]{article}
\usepackage[pdftex]{color,graphicx}
\usepackage[margin=0.8in]{geometry}
\usepackage[english]{babel}
\usepackage[justification=justified,font=small,labelfont=small]{caption}
\usepackage{subcaption}
\usepackage{appendix, parskip, inputenc, tocbibind, url}
\usepackage{amsmath, amssymb, amsthm, braket, empheq, fixltx2e, commath, mathtools}
\usepackage{cleveref}
\usepackage{ragged2e}
\usepackage{gensymb}
\usepackage{float}
\usepackage{cite}
\usepackage[ddmmyyyy]{datetime}

\begin{document}
\pagestyle{plain}
\setlength{\baselineskip}{14.4pt}


\begin{titlepage}
\begin{center}

\vspace*{60pt}
\LARGE{\textbf{Sub-picosecond proton tunnelling in deformed DNA hydrogen bonds under an asymmetric double-oscillator model}}

\vspace{14.4pt}
\large{\today}

\vspace{60pt}
\large{\textsc{J. Luo}*}

\vfill
\large{*galane.j.luo@gmail.com}

\vspace{14.4pt}
\large{Department of Mathematical Sciences, Durham University \\ Durham, DH1 3LE, United Kingdom}

\end{center}
\end{titlepage}


\begin{titlepage}
\begin{center}

\begin{minipage}{\textwidth}

\vspace*{120pt}
\begin{flushleft}
\Large{\textsc{Abstract}}
\end{flushleft}

\vspace{14.4pt}

We present a model of proton tunnelling across DNA hydrogen bonds, compute the characteristic tunnelling time (CTT) from donor to acceptor and discuss its biological implications. The model is a double oscillator characterised by three geometry parameters describing planar deformations of the H bond, and a symmetry parameter representing the energy ratio between ground states in the individual oscillators. We discover that some values of the symmetry parameter lead to CTTs which are up to 40 orders of magnitude smaller than a previous model predicted. Indeed, if the symmetry parameter is sufficiently far from its extremal values of 1 or 0, then the proton's CTT under any physically realistic planar deformation is guaranteed to be below one picosecond, which is a biologically relevant time-scale. This supports theories of links between proton tunnelling and biological processes such as spontaneous mutation. 

\end{minipage}
\end{center}
\end{titlepage}


	\section{Introduction} \label{intro}

In the DNA double helix, the two strands of nucleobases are held together by hydrogen bonds, each consisting of a proton being covalently bonded with a donor atom from a donor molecule, and electrostatically attracted to an acceptor atom from an acceptor molecule \cite{Pauling1960,Arunan2011}. L\"{o}wdin proposed that the proton in an H bond may break away from the donor atom and form a new covalent bond with the acceptor atom, by the mechanism of quantum tunnelling across the potential barrier between the donor and acceptor, and that this process may cause spontaneous mutation \cite{Lowdin1963}. McFadden and Al-Khalili later demonstrated that quantum coherence between the tunnelling proton and its environment can be maintained for biological time-scales, which validates modelling the proton's dynamics as being entirely quantum mechanical \cite{McFadden1999}.

In a \emph{normal} H bond, all atoms in the donor and acceptor molecules are co-planar, and the donor and acceptor atoms are co-linear with the proton. A planar deformation of the normal H bond is some combination of translations and rotations in the donor-acceptor molecular plane \cite{Dickerson1989,Lu1999,Olson2001}. It has been theorised that planar deformations of the H bond can have significant effects on the characteristic time-scale of proton tunnelling, and Krasilnikov studied these effects by modelling the potential in the H bond as a double harmonic oscillator which, when the bond is normal, is symmetric about the potential barrier \cite{Krasilnikov2014}. It was found that the characteristic tunnelling time (CTT) of the proton was extremely sensitive to bond deformation, taking values up to $\mathcal{O} (10^{27})$s, which was not a biologically relevant time-scale.  

We propose a generalisation to Krasilnikov's model, in which we associate the symmetry of the double-well potential in a normal H bond with a parameter, $\gamma$, whose value equals the energy ratio between a ground-state proton covalently bonded with the donor and one covalently bonded with the acceptor. When $\gamma$ takes its maximum value of 1, we recover Krasilnikov's model; when $0 < \gamma < 1$, the two local wells in the H bond potential are not equivalent, and the proton has a preferred equilibrium state near the donor rather than acceptor. We further encode the planar deformation of the H bond in three other parameters, $d_x, d_y$ representing relative shifts between the donor and acceptor, and $\theta$ representing the relative rotation, all of which are defined in detail in \Cref{model}. We then derive an analytical expression for the proton's CTT. Fixing all other parameters such as proton mass and covalent bond lengths at values appropriate to DNA H bonds, the CTT is a function of $\gamma, d_x, d_y$ and $\theta$. We discover that moderate values of $\gamma$ guarantee sub-picosecond proton tunnelling, regardless of bond deformation. In \Cref{conclusions}, we discuss the biological implications of our results.

	\section{Model and Results} \label{model}

In this Section, we firstly describe the geometry of an H bond under planar deformation, then define our double-well potential within this H bond, before solving the Schr\"{o}dinger equation under this potential to obtain the proton's wavefunction. From this wavefunction, we derive the proton's CTT. We make the following assumptions and approximations in our model. Firstly, we consider only \emph{stationary} bonds, meaning that the bond is not actively undergoing deformation whilst proton dynamics is taking place. Secondly, we assume that the lengths and relative angles of all \emph{covalent} bonds in the donor and acceptor molecules are unaffected by the deformation. In other words, we only consider translations and rotations of the donor molecule as a whole and, independently, of the acceptor molecule as a whole. Finally, even though the proton's global equilibrium is in a covalent bond with the donor atom, we assume that the proton can exist with a higher energy in a locally-stable state of being covalently bonded to the acceptor atom. That there are two local potential minima for the proton in the H bond is the foundation of our double-oscillator model. 

	\begin{figure}[ht]
	\centering
	\scalebox{0.64}{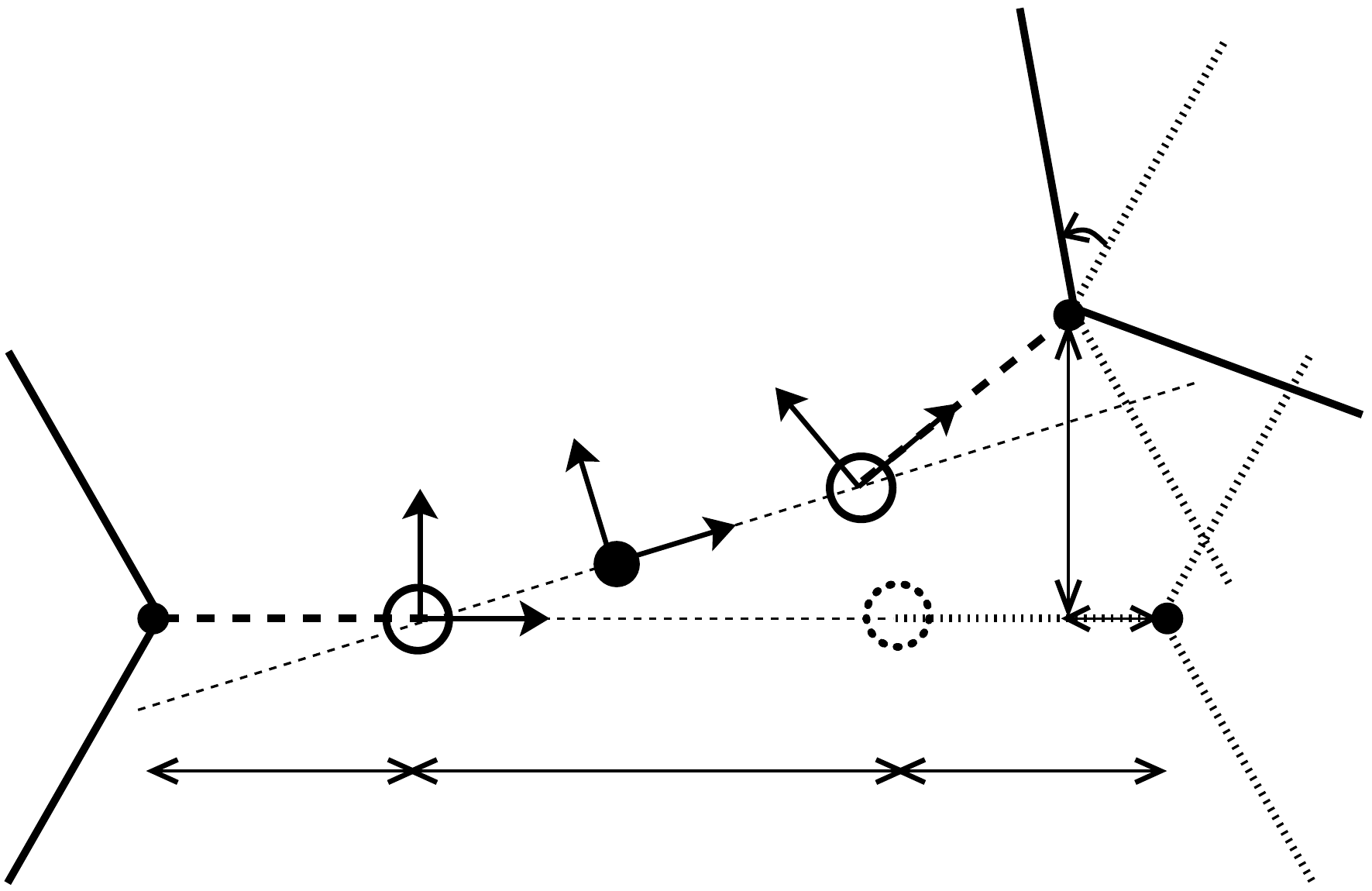}
	\captionsetup{width=0.64\textwidth}
	\caption{Geometry of a DNA H bond under planar deformation.} \label{figure01}
	\end{figure}

Since the H bond is planar, it suffices to model the potential for the proton as a function of two spatial dimensions. The geometry of the deformed H bond is shown in \Cref{figure01}. Thick lines marked A and D represent, respectively, the acceptor and donor molecules in a deformed bond, whilst the dotted line D$'$ marks where the donor molecule would be in a normal bond. N\textsubscript{1} and N\textsubscript{2} mark the acceptor and donor atoms, respectively, and N\textsubscript{2}$'$ marks where the donor atom would be in a normal bond. We set up three Cartesian coordinate systems as follows. Firstly, centred at O\textsubscript{1}, where a proton could exist in a covalent bond with N\textsubscript{1}, we have $(x_1,y_1)$, with $x_1$ pointing in the $\overrightarrow{\textnormal{N}_1 \textnormal{O}_1}$ direction. Secondly, centred at O\textsubscript{2}, where a proton could exist in a covalent bond with N\textsubscript{2}, we have $(x_2, y_2)$, with $x_2$ pointing  in the $\overrightarrow{\textnormal{O}_2 \textnormal{N}_2}$ direction. Lastly, centred at O, the saddle point in the double-well potential of the H bond, whose exact position along $\overrightarrow{\textnormal{O}_1 \textnormal{O}_2}$ depends upon our potential function, we have $(x,y)$, with $x$ pointing in the $\overrightarrow{\textnormal{O}_1 \textnormal{O}_2}$ direction. O\textsubscript{2}$'$ marks where O\textsubscript{2} would be in a normal bond. The bond geometry is entirely characterised by 5 parameters, which are marked in \Cref{figure01} as $L, D_0, d_x, d_y, \theta$, and defined as follows. $L$ is the distance between N\textsubscript{1} and O\textsubscript{1}, which we assume to be the same as the distance between N\textsubscript{2} and O\textsubscript{2}, as well as the distance bweteen N\textsubscript{2}$'$ and O\textsubscript{2}$'$, since we have assumed that no deformation affects the lengths of covalent bonds. $D_0$ is the distance between O\textsubscript{1} and O\textsubscript{2}$'$, in a normal bond. $d_x$ and $d_y$ are, respectively, the shifts in the $x_1$ and $y_1$ directions of the donor molecule from its normal position, so that, for instance, $d_x < 0$ represents a shift of the donor molecule \emph{towards} the acceptor molecule. Finally, $\theta$ is the anticlockwise angle by which the donor molecule is rotated from its normal orientation, about the point N\textsubscript{2}. We emphasise that the shifts are independent from the rotation, which means that the order in which $d_x, d_y$ and $\theta$ act on the system does not affect its final configuration. 

By comparing the coordinates of an arbitrary point in the three systems, O\textsubscript{1}$x_1 y_1$, O\textsubscript{2}$x_2 y_2$ and O$x y$, we write down the following coordinate transformation equations.
	\begin{subequations} \label{coordtrans}
	\begin{align}
	x_1 &= \left( x + \lambda D_{\theta} \right) \cos\theta_1 - y \sin\theta_1, \\
	y_1 &= \left( x + \lambda D_{\theta} \right) \sin\theta_1 + y \cos\theta_1, \\
	x_2 &= \left( x - ( 1 - \lambda ) D_{\theta} \right) \cos\theta_2 - y \sin\theta_2, \\
	y_2 &= \left( x - ( 1 - \lambda ) D_{\theta} \right) \sin\theta_2 + y \cos\theta_2,
	\end{align}
	\end{subequations}
where $\theta_1$ is the anticlockwise angle from $x_1$ to $x$, $\theta_2$ is the anticlockwise angle from $x_2$ to $x$, $D_{\theta}$ is the distance between $O_1$ and $O_2$ in the deformed bond, and $\lambda D_{\theta}$ where $0 < \lambda < 1$ is the distance between $O_1$ and $O$ in the deformed bond. We express $\theta_1, \theta_2, D_{\theta}$ and $\lambda$ in terms of $L, D_0, d_x, d_y$ and $\theta$ as follows.
	\begin{subequations} \label{coordtrans2}
	\begin{align}
	\theta &= 2\pi + \theta_1 - \theta_2, \\
	D_{\theta} \cos\theta_1 &= D_0 + L + d_x - L \cos \theta, \\
	D_{\theta} \sin\theta_1 &= d_y - L \sin \theta,
	\end{align}
	\end{subequations}
which imply
	\begin{subequations} \label{coordtrans3}
	\begin{align}
	D_{\theta} &= \sqrt{ \left[ D_0 + d_x + L \left( 1 - \cos \theta \right) \right]^2 + \left[ d_y - L \sin \theta \right]^2 } \\
	\cos\theta_1 &= \frac{D_{\theta}}{D_0 + d_x + L \left( 1 - \cos \theta \right)}, \quad \sin\theta_1 = \frac{D_{\theta}}{d_y - L \sin \theta}, \\
	\cos\theta_2 &= \cos\theta_1 \cos \theta + \sin\theta_1 \sin \theta, \quad \sin\theta_2 = \sin\theta_1 \cos \theta - \cos\theta_1 \sin \theta,
	\end{align}
	\end{subequations}
and $\lambda$ is dependent upon the form of the potential function over the $(x,y)$ plane. For our asymmetric double-oscillator model, we consider a potential function $V = V_1 + V_2$, with
	\begin{subequations} \label{potential1}
	\begin{align}
	V_1 ( x , y ) &= \Big\{ \begin{array}{ll} U_1 ( x_1 , y_1 ) := \frac{1}{2} m \omega_1^2 \left( x_1^2 + g^2 y_1^2 \right) & \quad \textnormal{if}~ - \infty < x < 0, ~ - \infty < y < \infty \\ 0 & \quad \textnormal{otherwise} \end{array} , \label{U1x1y1}  \\
	V_2 ( x , y ) &= \Big\{ \begin{array}{ll} U_2( x_2 , y_2 ) := \frac{1}{2} m \omega_2^2 \left( x_2^2 + g^2 y_2^2 \right) & \quad \textnormal{if}~ 0 \le x < \infty, ~ - \infty < y < \infty \\ 0 & \quad \textnormal{otherwise} \end{array} , \label{U2x2y2}
	\end{align}
	\end{subequations}
and the proton wavefunction, $\Psi(x,y,t)$, evolves in time according to the Schr\"{o}dinger equation, 
	\begin{align}
	i \hbar \frac{\textnormal{d} \Psi}{\textnormal{d} t} = \left[ - \frac{\hbar^2}{2m} \left( \partial_x^2 + \partial_y^2 \right) + V \right] \Psi. \label{SE}
	\end{align}
In \cref{potential1,SE}, $m$ is proton mass, $\omega_1$ and $\omega_2$ respectively are natural angular frequencies of the single oscillators $U_1$ and $U_2$, and $g > 0$ is an \emph{isotropy parameter} which we assume to be the same for $U_1$ and $U_2$. We define the \emph{symmetry parameter}, 
	\begin{align}
	\gamma := \omega_2 / \omega_1 \le 1, \label{defn_gamma}
	\end{align}
so that if $\gamma < 1$ then there is a lower ground state in $U_2$ than in $U_1$, and this represents the fact that the proton's preferred equilibrium is in $U_2$. $V$ is a double oscillator which is identical to $U_1$ to the left of the line $x = 0$ and identical to $U_2$ to the right of $x = 0$. Thus, there is a potential barrier along the line $x = 0$ where, in general, we have $U_1 \neq U_2$, so that there is a discontinuity in $V$. 

With the potential function in place, we now calculate $\lambda$. In the O\textsubscript{1}$x_1 y_1$ frame, the local potential well's equipotential curve through the point O is an ellipse, with equation $x_1^2 + g^2 y_1^2 = 2 U_0 / ( m \omega_1^2 )$, where $U_0$ is the potential energy at O. One could write a similar ellipse equation, in terms of $(x_2,y_2)$, for the equipotential curve through O in $U_2$. Instead, using \cref{coordtrans,coordtrans2}, we write both ellipse equations in the O$xy$ frame, as follows. 
	\begin{subequations} \label{ellipses}
	\begin{align}
	\left[ \left( x + \lambda D_{\theta} \right) \cos\theta_1 - y \sin\theta_1 \right]^2 + g^2 \left[ \left( x + \lambda D_{\theta} \right) \sin\theta_1 + y \cos\theta_1 \right]^2 &= \frac{ 2 U_0 }{ m \omega_1^2 }, \label{ellipse1} \\
	\left[ \left( x - ( 1 - \lambda ) D_{\theta} \right) \cos\theta_2 - y \sin\theta_2 \right]^2 + g^2 \left[ \left( x - ( 1 - \lambda ) D_{\theta} \right) \sin\theta_2 + y \cos\theta_2 \right]^2 &= \frac{ 2 U_0 }{ m \omega_2^2 }. \label{ellipse2}
	\end{align}
	\end{subequations}
Since the ellipses intersect at O, we set $(x,y) = (0,0)$ in \cref{ellipse1,ellipse2}, to obtain 
	\begin{align}
	U_0 = \frac{m \omega_1^2}{2} \lambda^2 D_{\theta}^2 \left( \cos^2\theta_1 + g^2 \sin^2\theta_1 \right) = \frac{m \omega_2^2}{2} ( 1 - \lambda )^2 D_{\theta}^2 \left( \cos^2\theta_2 + g^2 \sin^2\theta_2 \right), \label{U0}
	\end{align}
from which it follows that
	\begin{align}
	\lambda = \left( 1 + \frac{1}{\gamma} \sqrt{ \frac{ \cos^2\theta_1 + g^2 \sin^2\theta_1 }{ \cos^2\theta_2 + g^2 \sin^2\theta_2 } } \right)^{-1}. \label{lambda}
	\end{align}
We proceed to compute the characteristic time-scale of proton tunnelling from being localised in $U_2$ to being maximally localised in $U_1$, using the Rayleigh-Ritz ansatz \cite{Merzbacher1998}, in which the ground state wavefunction of the proton is approximately
	\begin{align}
	\Psi (x,y,t) = \alpha_1 (t) \phi_1 (x,y) + \alpha_2 (t) \phi_2 (x,y), \label{defn_Ps}
	\end{align}
where $\alpha_{1,2}$ are complex coefficients, and $\phi_{1,2}$ are normalised ground state wavefunctions that the proton would have if it existed in the single-well potential $U_1$ or $U_2$, with their domains extended to the infinite plane. We note that if a proton were in the single oscillator $U_1$ or $U_2$, then its ground state energy would be
	\begin{align}
	E_1 := \hbar \omega_1 (1+g) / 2 ~\textnormal{for}~U_1\quad \textnormal{or} \quad E_2 := \hbar \omega_2 (1+g) / 2 ~\textnormal{for}~U_2, \label{E1E2}
	\end{align} 
so that the symmetry parameter, $\gamma$, equals the energy ratio $E_2 / E_1$. Scaling length by
	\begin{align}
	x_0 := \sqrt{ \frac{ \hbar }{m \omega_1} }, \label{defn_x0}
	\end{align}
we have $\phi_1$ and $\phi_2$ in the following dimensionless forms, in terms of coordinates $\xi_{1,2} := x_{1,2} / x_0$ and $\eta_{1,2} := y_{1,2} / x_0$.
	\begin{subequations} \label{ph1ph2}
	\begin{align}
	\phi_1 ( \xi_1, \eta_1 ) &= \frac{ g^{1/4} }{ \sqrt{\pi} } \exp \left[ - \frac{1}{2} \left( \xi_1^2 + g \eta_1^2 \right) \right], \quad - \infty < \xi_1, \eta_1 < \infty, \\
	\phi_2 ( \xi_2, \eta_2 ) &= \frac{ g^{1/4} \sqrt{\gamma} }{ \sqrt{\pi} } \exp \left[ - \frac{\gamma}{2} \left( \xi_2^2 + g \eta_2^2 \right) \right], \quad - \infty < \xi_2, \eta_2 < \infty.
	\end{align}
	\end{subequations}
Scaling time by $\omega_1^{-1}$, then $\Psi$ evolves according to the dimensionless Schr\"{o}dinger equation,
	\begin{align}
	i \frac{ \textnormal{d} \Psi}{ \textnormal{d} \tau} = \widehat{H} \Psi, \label{dimlessSE}
	\end{align}
where $\tau$ is dimensionless time and, in coordinates $(\xi, \eta) = (x, y) / x_0$, we have
	\begin{align}
	\widehat{H} = \frac{1}{\hbar \omega_1} \left[ - \frac{ \hbar^2 }{ 2 m } \left( \partial_x^2 + \partial_y^2 \right) + V \right] = - \frac{1}{2} \left( \partial_\xi^2 + \partial_\eta^2 \right) + v_1 + v_2,
	\end{align}
with
	\begin{subequations} \label{v1v2}
	\begin{align}
	v_1 ( \xi, \eta ) &= \Big\{ \begin{array}{ll} u_1 (\xi_1 , \eta_1 ) := \frac{1}{2} \left( \xi_1^2 + g^2 \eta_1^2 \right) & \quad  \textnormal{if}~ - \infty < \xi < 0, ~ - \infty < \eta < \infty \\ 0 & \quad \textnormal{otherwise} \end{array} , \label{v1} \\
	v_2 ( \xi, \eta ) &= \Big\{ \begin{array}{ll} u_2 (\xi_2 , \eta_2 ) := \frac{\gamma^2}{2} \left( \xi_2^2 + g^2 \eta_2^2 \right) & \quad \textnormal{if}~ 0 \le \xi < \infty, ~ - \infty < \eta < \infty \\ 0 & \quad \textnormal{otherwise} \end{array} . \label{v2}
	\end{align}
	\end{subequations}
Since $\partial_\xi^2 + \partial_\eta^2 = \partial_{\xi_{1,2}}^2 + \partial_{\eta_{1,2}}^2$, we have the following identities.
	\begin{align}
	\widehat{H} \phi_1 = \left( \frac{1}{2} ( 1 + g ) - u_1 + v_1 + v_2 \right) \phi_1, \quad \widehat{H} \phi_2 = \left( \frac{\gamma}{2} ( 1 + g ) - u_2 + v_1 + v_2 \right) \phi_2, \label{hHph}
	\end{align}
where $u_{1,2}$ and $\phi_{1,2}$ are expressed in terms of coordinates $(\xi, \eta)$ as follows. Defining
	\begin{align}
	\Delta_{\theta} := D_{\theta} / x_0, \label{defn_Deth}
	\end{align}
and using the dimensionless version of \cref{coordtrans}, we obtain, for $j = 1, 2$,
	\begin{align}
	u_j = \frac{1}{2} \left( a_j \xi^2 + b_j \eta^2 + 2 c_j \xi \eta + 2 p_j \xi + 2 q_j \eta + r_j \right),
	\end{align}	
where
	\begin{subequations} \label{defn_a1}
	\begin{align}
	a_1 &= \cos^2\theta_1 + g^2 \sin^2\theta_1, \quad a_2 = \gamma^2 \left( \cos^2\theta_2 + g^2 \sin^2\theta_2 \right), \\
	b_1 &= \sin^2\theta_1 + g^2 \cos^2\theta_1, \quad b_2 = \gamma^2 \left( \sin^2\theta_2 + g^2 \cos^2\theta_2 \right), \\
	c_1 &= \left( g^2 - 1 \right) \cos\theta_1 \sin\theta_1, \quad c_2 = \gamma^2 \left( g^2 - 1 \right) \cos\theta_2 \sin\theta_2, \\
	p_1 &= a_1 \lambda \Delta_{\theta}, \quad p_2 = - a_2 \left( 1 - \lambda \right) \Delta_{\theta}, \\
	q_1 &= c_1 \lambda \Delta_{\theta}, \quad q_2 = - c_2 \left( 1 - \lambda \right) \Delta_{\theta}, \\
	r_1 &= a_1 \lambda^2 \Delta_{\theta}^2, \quad r_2 = a_2 \left( 1 - \lambda \right)^2 \Delta_{\theta}^2.
	\end{align}
	\end{subequations}
We note that $\lambda$ [cf. \cref{lambda}] can now be written
	\begin{align}
	\lambda = \left( 1 + \sqrt{ a_1 / a_2 } \right)^{-1},
	\end{align}
from which it follows that $r_1 = r_2$. We therefore define
	\begin{align}
	r_0 := r_1 = r_2 = \frac{a_1 a_2 \Delta_{\theta}^2}{\left( \sqrt{a_1} + \sqrt{a_2} \right)^2}. \label{defn_r0}
	\end{align}
For $\phi_j$ with $j = 1,2$, we have, for $-\infty < \xi, \eta < \infty$, 
	\begin{align}
	\phi_j ( \xi, \eta ) = \frac{ g^{1/4} }{ \sqrt{\pi} } \gamma^{ \frac{ j - 1 }{2} }\exp \left[ - \frac{1}{2} \left( A_j \xi^2 + B_j \eta^2 + 2 C_j \xi \eta + 2 P_j \xi + 2 Q_j \eta + R_j \right) \right],
	\end{align}
where 
	\begin{subequations}
	\begin{align}
	A_1 &= \cos^2\theta_1 + g \sin^2\theta_1, \quad A_2 = \gamma \left( \cos^2\theta_2 + g \sin^2\theta_2 \right), \\
	B_1 &= \sin^2\theta_1 + g \cos^2\theta_1, \quad B_2 = \gamma \left( \sin^2\theta_2 + g \cos^2\theta_2 \right), \\
	C_1 &= \left( g - 1 \right) \cos\theta_1 \sin\theta_1, \quad C_2 = \gamma \left( g - 1 \right) \cos\theta_2 \sin\theta_2, \\
	P_1 &= A_1 \lambda \Delta_{\theta}, \quad P_2 = - A_2 \left( 1 - \lambda \right) \Delta_{\theta}, \\
	Q_1 &= C_1 \lambda \Delta_{\theta}, \quad Q_2 = - C_2 \left( 1 - \lambda \right) \Delta_{\theta}, \\
	R_1 &= A_1 \lambda^2 \Delta_{\theta}^2, \quad R_2 = A_2 \left( 1 - \lambda \right)^2 \Delta_{\theta}^2.
	\end{align}
	\end{subequations}
Defining the inner product $\braket{f|g} := \int_{-\infty}^{\infty} \textnormal{d} \xi \int_{-\infty}^{\infty} \textnormal{d} \eta~ f^* g$, we take the inner product of \cref{dimlessSE} with $\bra{\phi_1}$ and $\bra{\phi_2}$ respectively to obtain
	\begin{align}
	i\left(\begin{array}{cc} 1 & S \\ S & 1 \end{array} \right) \left(\begin{array}{c} \dot{\alpha}_1 \\ \dot{\alpha}_2\end{array} \right) = \left(\begin{array}{cc} H_{11} & H_{12} \\ H_{21} & H_{22} \end{array} \right) \left(\begin{array}{c} \alpha_1 \\ \alpha_2 \end{array} \right) , \label{lineareqn}
	\end{align}
where the overdot denotes differentiation with respect to $\tau$, and
	\begin{align}
	S = \braket{\phi_1 | \phi_2}, \quad H_{jk} = \braket{\phi_j | \widehat{H} \phi_k}. \label{defn_integrals}
	\end{align}
We note that since $\phi_1, \phi_2$ are positve, square normalised functions, and since $\phi_1 \not\equiv \phi_2$, we have $0 < S < 1$. Next, using \cref{hHph}, we deduce
	\begin{align}
	\left(\begin{array}{cc} H_{11} & H_{12} \\ H_{21} & H_{22} \end{array} \right) = \left(\begin{array}{cc} \frac{1}{2} \left( 1 + g \right) + I_{11} & \frac{\gamma}{2} \left( 1 + g \right) S - I_{12} \\ \frac{1}{2} \left( 1 + g \right) S + I_{21} & \frac{\gamma}{2} \left( 1 + g \right) - I_{22} \end{array}\right) , \label{Hij} 
	\end{align}
where
	\begin{align}
	\left(\begin{array}{cc} I_{11} & I_{12} \\ I_{21} & I_{22} \end{array} \right) = \left(\begin{array}{cc} \int_0^{\infty} \textnormal{d} \xi \int_{-\infty}^{\infty} \textnormal{d} \eta \left( u_2 - u_1 \right) \phi_1^2 & \int_{-\infty}^0 \textnormal{d} \xi \int_{-\infty}^{\infty} \textnormal{d} \eta \left( u_2 - u_1 \right) \phi_1 \phi_2 \\ \int_0^{\infty} \textnormal{d} \xi \int_{-\infty}^{\infty} \textnormal{d} \eta \left( u_2 - u_1 \right) \phi_1 \phi_2 & \int_{-\infty}^0 \textnormal{d} \xi \int_{-\infty}^{\infty} \textnormal{d} \eta \left( u_2 - u_1 \right) \phi_2^2 \end{array}\right), \label{Iij}
	\end{align}
By invoking the change of variable $\xi \mapsto - \xi$ where necessary, we write, for $j = 1,2$ and $k = 1,2$,
	\begin{align} \label{Iij_details}
	I_{jk} &= \frac{ \sqrt{g} }{2 \pi} \gamma^{ \frac{ j + k }{2} - 1} \int_0^{\infty} \textnormal{d} \xi \int_{-\infty}^{\infty} \textnormal{d} \eta \left( a \xi^2 + b \eta^2 + 2 (-1)^{k-1} c \xi \eta + 2 (-1)^{k-1} p \xi + 2 q \eta \right) \nonumber \\
	&\qquad \exp \left[ - \frac{1}{2} \left( A_{jk} \xi^2 + B_{jk} \eta^2 + 2 (-1)^{k-1} C_{jk} \xi \eta + 2 (-1)^{k-1} P_{jk} \xi + 2 Q_{jk} \eta + R_{jk} \right) \right],
	\end{align}
where $a = a_2 - a_1, A_{jk} = A_j + A_k$, and analogous definitions hold for $b, B_{jk}, c, C_{jk}, p, P_{jk}, q, Q_{jk}$ and $R_{jk}$. Each \emph{transition integral} $I_{jk}$ can be evaluated exactly, as can the \emph{overlap integral}, $S$. We present closed-form expressions for these integrals in the Appendix. 

To solve \cref{lineareqn} for $\alpha_j (\tau)$, we write
	\begin{align}
	\left(\begin{array}{c} \dot{\alpha}_1 \\ \dot{\alpha}_2 \end{array} \right) = \left( \begin{array}{cc} J & K \\ M & N \end{array} \right) \left(\begin{array}{c} \alpha_1 \\ \alpha_2 \end{array} \right) , \label{lineareqn2}
	\end{align}
where
	\begin{align}
	\left( \begin{array}{cc} J & K \\ M & N \end{array} \right) &= \frac{-i}{1 - S^2}\left( \begin{array}{cc} 1 & -S \\ -S & 1 \end{array} \right) \left(\begin{array}{cc} H_{11} & H_{12} \\ H_{21} & H_{22} \end{array} \right) \nonumber \\
	&= -i \left( \begin{array}{cc} \frac{1}{2} \left( 1 + g \right) + \frac{ I_{11} - S I_{21} }{1 - S^2} & - \frac{ I_{12} - S I_{22} }{1 - S^2} \\ \frac{ I_{21} - S I_{11} }{1 - S^2 } & \frac{\gamma}{2} \left( 1 + g \right) - \frac{ I_{22} - S I_{12}}{1 - S^2 } \end{array} \right). \label{lineareqn3} 
	\end{align}
The solution of \cref{lineareqn2} subject to the initial condition, $(\alpha_1, \alpha_2) = (0,1)$ at $\tau = 0$, is
	\begin{align}
	\Big( \begin{array}{c} \alpha_1 \\ \alpha_2 \end{array} \Big) = \frac{1}{\mathcal{N}_\tau} \left( \beta_+\boldsymbol{r}_+ e^{\tau \rho_+} + \beta_- \boldsymbol{r}_- e^{\tau \rho_-} \right), \label{solution1}
	\end{align}
where
	\begin{align}
	\rho_{\pm} = \frac{ J + N \pm \Omega }{2}, \quad \boldsymbol{r}_{\pm} = \left( 1, \frac{ - J + N \pm \Omega }{ 2 K } \right)^T, \quad \beta_{\pm} = \pm K / \Omega, \label{solution2}
	\end{align}
with
	\begin{align}
	\Omega = \sqrt{ ( J - N )^2 + 4 K M }, \label{defn_O}
	\end{align}
and we determine the real function $\mathcal{N}_\tau$ as follows. From \cref{solution1}, we have
	\begin{subequations} \label{a1a2_more}
	\begin{align}
	\alpha_1 &= \frac{2 K}{\mathcal{N}_\tau \Omega} \exp \left( \frac{ J + N }{2} \tau \right) \sinh \frac{ \Omega \tau }{2} , \\
	\alpha_2 &= \frac{1}{\mathcal{N}_\tau} \exp\left(\frac{J+N}{2}\tau \right)\left[\cosh\frac{\Omega\tau}{2} - \frac{(J-N)}{\Omega} \sinh\frac{\Omega\tau}{2} \right].
	\end{align}
	\end{subequations}
Assume for now that $\Omega^2 < 0$, which we later verify numerically, so that $\Omega = i | \Omega | $, then we have
	\begin{align}
	\sinh \frac{ \Omega \tau}{2} = i \sin \frac{ |\Omega| \tau }{2}, \quad \cosh \frac{ \Omega \tau}{2} = \cos \frac{ |\Omega| \tau }{2}. \label{sinh_cosh}
	\end{align}
Since $| e^{(J+N)\tau /2} | = 1$, it follows from the normalisation condition, $\braket{\Psi | \Psi} = | \alpha_1 |^2 + | \alpha_2 |^2 + ( \alpha_1^* \alpha_2 + \alpha_2^* \alpha_1 ) S = 1$, that
	\begin{align}
	\mathcal{N}_\tau = \sqrt{\cos^2 \frac{ |\Omega| \tau }{2} + \sigma \sin^2 \frac{ |\Omega| \tau }{2}},
	\end{align}
where $\sigma = ( 4|K|^2 + |J-N|^2 + 4K(J-N)S ) / | \Omega |^2$, which is real because $K (J-N)$ is real. Since $S < 1$, we have $\sigma > ( 4|K|^2 + |J-N|^2 - 4 |K (J-N) | ) / | \Omega |^2 = ( 2 |K| - | J - N | )^2 / | \Omega |^2$, therefore $\sigma > 0$. In the proton wavefunction $\Psi = \alpha_1 \phi_1 + \alpha_2 \phi_2$, $\alpha_2$ is initially unity and $\alpha_1$ is initially zero, so we say that the proton's CTT, the time it takes for $\Psi$ to evolve from being localised as $\phi_2$ to being maximally localised in the potential well $u_1$, is the time at which 
	\begin{align}
	| \alpha_1 | = \frac{ 2 |K| } { \mathcal{N}_\tau |\Omega| } \abs{ \sin \frac{ | \Omega | \tau }{2} } 
	\end{align}
first reaches its maximum. This happens at the smallest $\tau$ for which the following holds.
	\begin{align}
	0 = \frac{\textnormal{d} }{\textnormal{d} \tau } \frac{\sin \frac{ | \Omega | \tau }{2} }{\mathcal{N}_\tau } &= \frac{ | \Omega | }{2 \mathcal{N}_\tau} \cos \frac{ |\Omega| \tau }{2} + \frac{| \Omega |}{ 2 \mathcal{N}_\tau^3 } \left( \cos \frac{ | \Omega | \tau }{2} \sin \frac{ | \Omega | \tau }{2} - \sigma \sin \frac{ | \Omega | \tau }{2} \cos \frac{ | \Omega | \tau }{2} \right) \sin \frac{ |\Omega| \tau }{2} \nonumber \\
	&= \frac{|\Omega|}{2 \mathcal{N}_\tau^3} \cos \frac{ |\Omega| \tau }{2}.
	\end{align}
Therefore, the CTT of the proton is $\tau_\textnormal{p} = \pi / | \Omega |$, or, in physical units,
	\begin{align}
	t_\textnormal{p} = \frac{\pi}{\omega_1 | \Omega |}, \label{CTT} 
	\end{align}
where, due to \cref{lineareqn3,defn_O}, we have
	\begin{align}
	\Omega = \sqrt{ - \left[ \frac{1}{2} \left( 1 + g \right) \left( 1 - \gamma \right) + \frac{I_{11} + I_{22} - S \left( I_{12} + I_{21} \right)}{1 - S^2} \right]^2+ \frac{ 4 \left( I_{12} - S I_{22} \right) \left( I_{21} - S I_{11} \right)}{\left( 1 - S^2 \right)^2} }.
	\end{align}

We have the following values for the parameters $D_0, L$ and $g$ which are appropriate for H bonds across the DNA double helix \cite{Ishenko1985,Santamaria1999,FonsecaGuerra2000,Steiner2002}. $4.5 \times 10^{14} \textnormal{s}^{-1} \le \omega_1 \le 6.4 \times 10^{14} \textnormal{s}^{-1}, 0.61 \textnormal{\r{A}} \le D_0 \le 0.81 \textnormal{\r{A}}, 1.03 \textnormal{\r{A}} \le L \le 1.07 \textnormal{\r{A}}, g \approx 0.5$. We fix $\omega_1 = 5.45 \times 10^{14} \textnormal{s}^{-1}, D_0 = 0.71 \textnormal{\r{A}}, L = 1.05 \textnormal{\r{A}}, g = 0.5$, and compute $t_\textnormal{p}$ as functions of the parameters $\gamma, d_x, d_y$ and $\theta$. For all parameter values which we have studied, we find $\Omega^2 < 0$, which ensures that \cref{sinh_cosh} holds. We note also that when $\gamma = 1$, we recover results of \cite{Krasilnikov2014} relating to deformations of a \emph{symmetric} double oscillator. 

In order for our model to represent tunnelling, rather than scattering, we must have the height $U_0$ [cf. \cref{U0}] of the saddle point in the double-well potential surface being greater than the ground-state energy of $\phi_2$ [cf. \cref{E1E2}]; that is, we must have
	\begin{align}
	u_0  := \frac{U_0}{E_2} = \frac{r_0}{\gamma (1+g)} > 1. \label{u0_cond}
	\end{align}
Moreover, the expected value of proton energy must be conserved by the tunnelling process; that is, we must have $\textnormal{d} \braket{ \Psi | \widehat{H} \Psi } / \textnormal{d} \tau = 0$. Since $\widehat{H}$ is time-independent, we do indeed have $\textnormal{d} \braket{ \Psi | \widehat{H} \Psi } / \textnormal{d} \tau = \braket{ \dot{ \Psi } | \widehat{H} \Psi} + \braket{ \Psi | \widehat{H} \dot{ \Psi } } = i \braket{ \Psi | \widehat{H} \widehat{H} \Psi } - i \braket{ \Psi | \widehat{H} \widehat{H} \Psi } = 0$, where we have made use of the Schr\"{o}dinger equation and its dual, $- i \bra{ \dot{ \Psi } } = \bra{ \Psi } \widehat{H} $. We note that the proton wavefunction for $\tau > 0$ is always a superposition of $\phi_1$ and $\phi_2$ with a non-zero coefficient for $\phi_2$ [cf. \cref{a1a2_more}], for if that coefficient were to vanish at any time then the proton energy at that time would equal $E_1 > E_2$, violating the energy conservation requirement.

The deformation parameters $d_x, d_y$ and $\theta$ are encoded in $r_0$, as per the definition of \cref{defn_r0}. Our results show that, for each value of $\gamma$, there exists some critical value $d_x^\textnormal{crit}$ such that, if $d_x \ge d_x^\textnormal{crit}$ then \cref{u0_cond} is satisfied given any combination of $(d_y,\theta)$, whereas if $d_x < d_x^\textnormal{crit}$ then there are some combinations of $(d_y,\theta)$ under which \cref{u0_cond} fails to hold.
	\begin{figure}
	\centering
	\begin{minipage}{.5\textwidth}
	\centering
	\includegraphics[width=.95\linewidth]{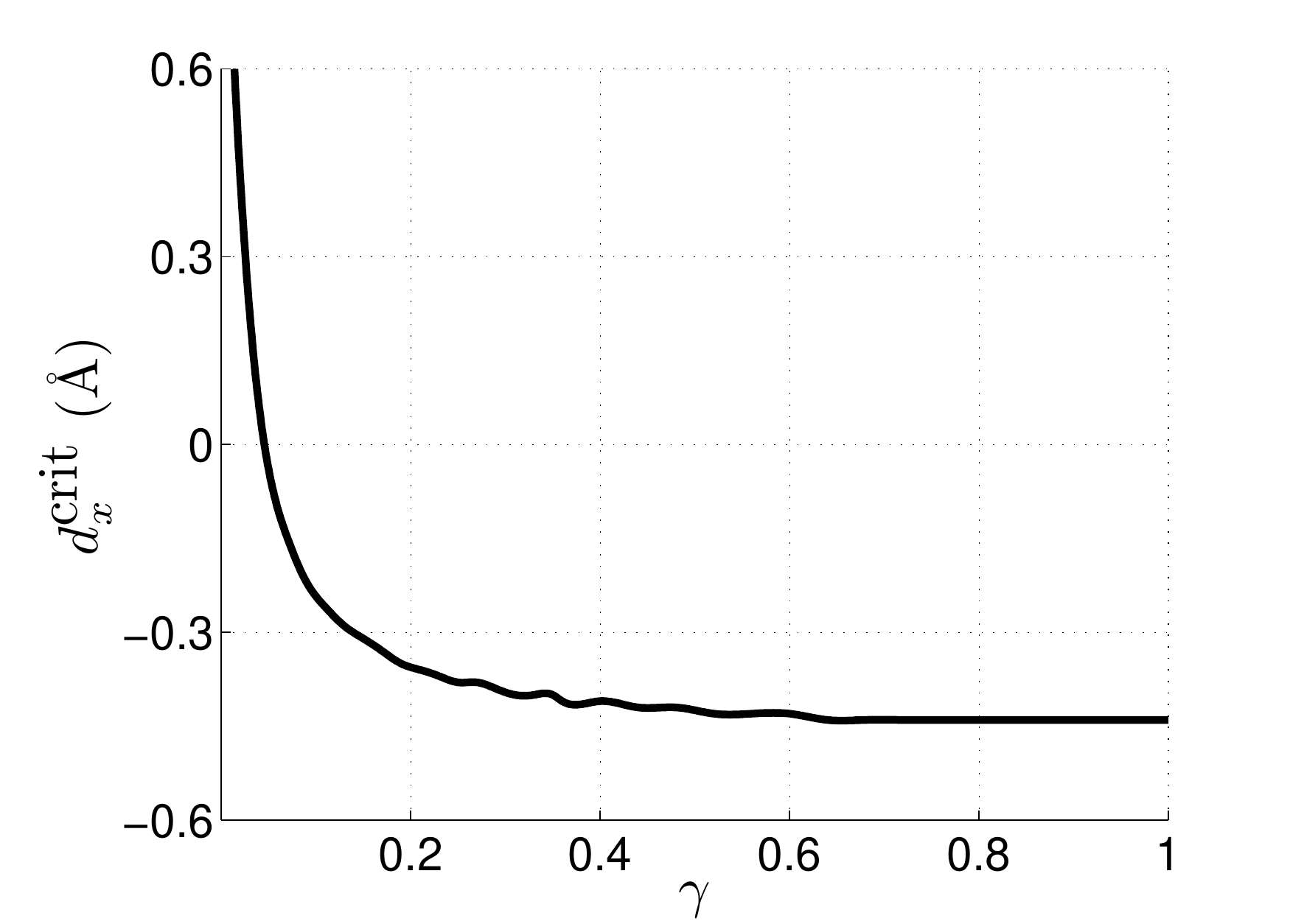}
	\caption{$d_x^\textnormal{crit}$ as a function of $\gamma$.}
	\label{figure02}
	\end{minipage}%
	\begin{minipage}{.5\textwidth}
	\centering
	\includegraphics[width=.95\linewidth]{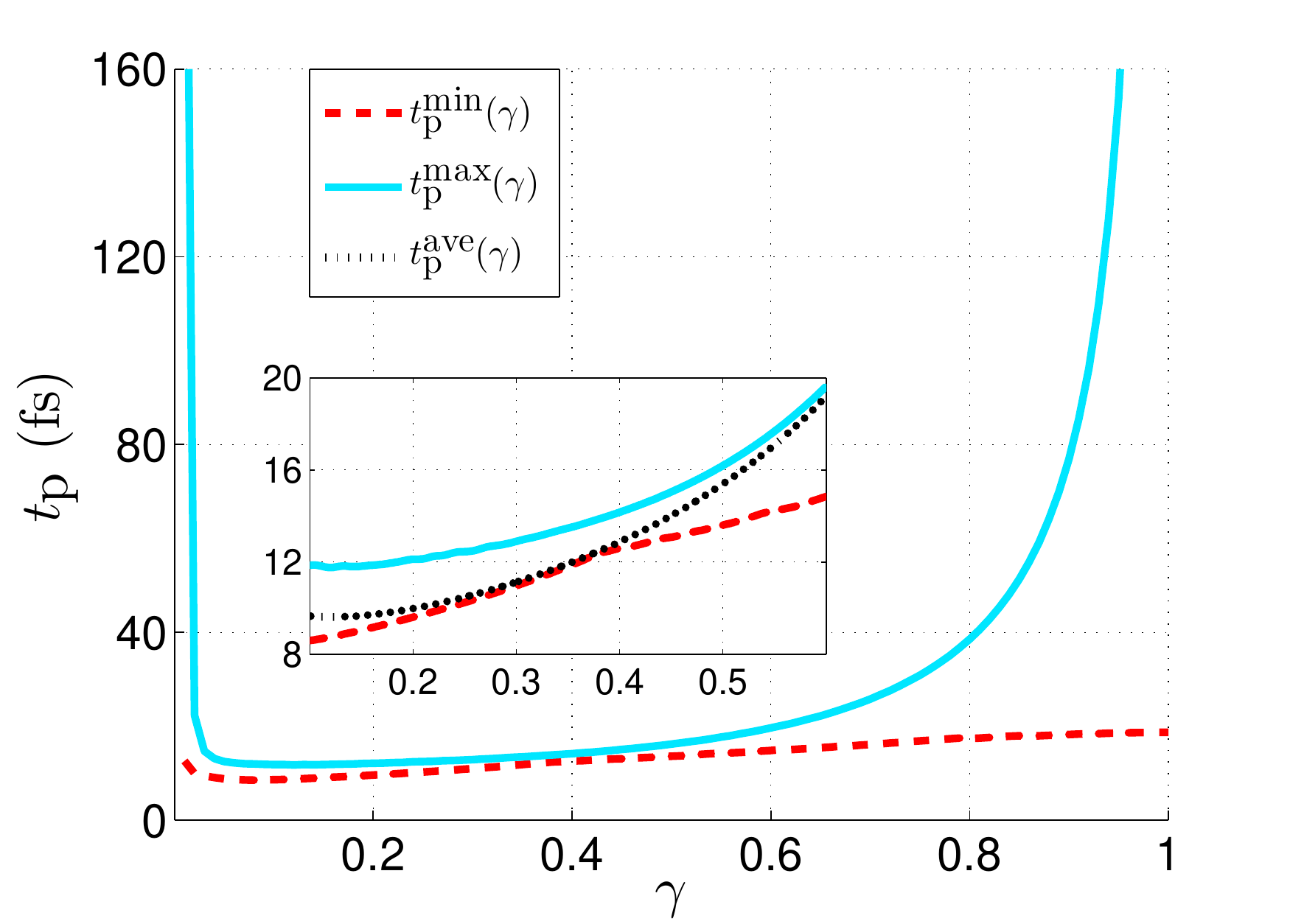}
	\caption{Min, max, average $t_\textnormal{p}$ as functions of $\gamma$.}
	\label{figure03}
	\end{minipage}
	\end{figure}
\Cref{figure02} shows $d_x^\textnormal{crit}$ as a function of $\gamma$. As $\gamma$ decreases towards 0, greater values of $d_x$ would be needed in order to guarantee that every combination of $(d_y,\theta)$ produces a valid tunnelling model. This is because $\gamma$ is positively correlated with the steepness of the local potential well $U_2 (x_2, y_2)$. The smaller $\gamma$ is, the further away from $(x_2,y_2) = (0,0)$ one needs to go before $U_2$ reaches the required height, namely the ground-state energy of $\phi_2$; thus, in order to ensure that the saddle point between $U_1$ and $U_2$ is sufficiently high, $U_1$ and $U_2$ must be far enough apart, hence the large $d_x^\textnormal{crit}$. Meanwhile, as $\gamma \rightarrow 1$, we observe that $d_x^\textnormal{crit} \rightarrow -0.44$\r{A}.

For $0.01 \le \gamma \le 1$, we vary $d_x, d_y, \theta$ as follows. $-0.45 \textnormal{\r{A}} \le d_x, d_y \le 0.45 \textnormal{\r{A}}, -90^\circ \le \theta \le 90^\circ$, and we only consider combinations of $(\gamma, d_x, d_y, \theta)$ such that \cref{u0_cond} holds. We find that for each $\gamma$, $t_\textnormal{p}$ falls in a range between some $t_\textnormal{p}^\textnormal{min} (\gamma)$ and some $t_\textnormal{p}^\textnormal{max} (\gamma)$, and in \Cref{figure03} we present these extremal values as functions of $\gamma$. Crucially, our results show that for $0.01 \le \gamma \le 0.99$, we always have $8.5\textnormal{fs} \le t_\textnormal{p} (\gamma, d_x, d_y, \theta) \le 770$fs. We also observe that $t_\textnormal{p}^\textnormal{max} (\gamma)$ increases steeply both as $\gamma \rightarrow 0$ and as $\gamma \rightarrow 1$. Indeed, when $\gamma = 1$, $t_\textnormal{p}^\textnormal{max} (\gamma)$ becomes $\sim \mathcal{O} (10^{27})$s; and even though $t_\textnormal{p}^\textnormal{min} (\gamma)$ is still $\sim \mathcal{O} (10^{-14})$s, $t_\textnormal{p}$ increases rapidly as $(d_x, d_y, \theta)$ moves away from the combination which minimises $t_\textnormal{p}$. Moreover, $t_\textnormal{p}^\textnormal{min} (\gamma)$ is slowly varying with $\gamma$, and there is a range of values of $\gamma$, namely $0.2 \lessapprox \gamma \lessapprox 0.4$, for which $t_\textnormal{p}^\textnormal{min} (\gamma)$ becomes close to $t_\textnormal{p}^\textnormal{max} (\gamma)$. In this case, varying $(d_x, d_y, \theta)$ has little effect on $t_\textnormal{p}$, which contrasts strongly with the large-$\gamma$ and small-$\gamma$ cases where $t_\textnormal{p}$ is very sensitive to $(d_x, d_y, \theta)$. We have defined $t_\textnormal{p}^\textnormal{ave} (\gamma)$ as the mean $t_\textnormal{p}$, given a fixed $\gamma$, over all combinations of $(d_x, d_y, \theta)$ which satisfy \cref{u0_cond}, and we have presented $t_\textnormal{p}^\textnormal{ave} (\gamma)$ for $0.1 \le \gamma \le 0.6$ in the small box in \Cref{figure03}. As $\gamma \rightarrow 1$, we have $t_\textnormal{p}^\textnormal{ave} (\gamma) \sim t_\textnormal{p}^\textnormal{max} (\gamma)$, and for intermediate values of $\gamma$,  namely $\gamma \approx 0.3$, we have $t_\textnormal{p}^\textnormal{ave} (\gamma) \sim t_\textnormal{p}^\textnormal{min} (\gamma)$, but as $\gamma \rightarrow 0$, $t_\textnormal{p}^\textnormal{ave} (\gamma)$ is asymptotic to neither $t_\textnormal{p}^\textnormal{min} (\gamma)$ nor $t_\textnormal{p}^\textnormal{max} (\gamma)$. 
	\begin{figure}[h!]
	\centering
	\begin{subfigure}{.85\textwidth}
	\centering
	\includegraphics[width=\linewidth]{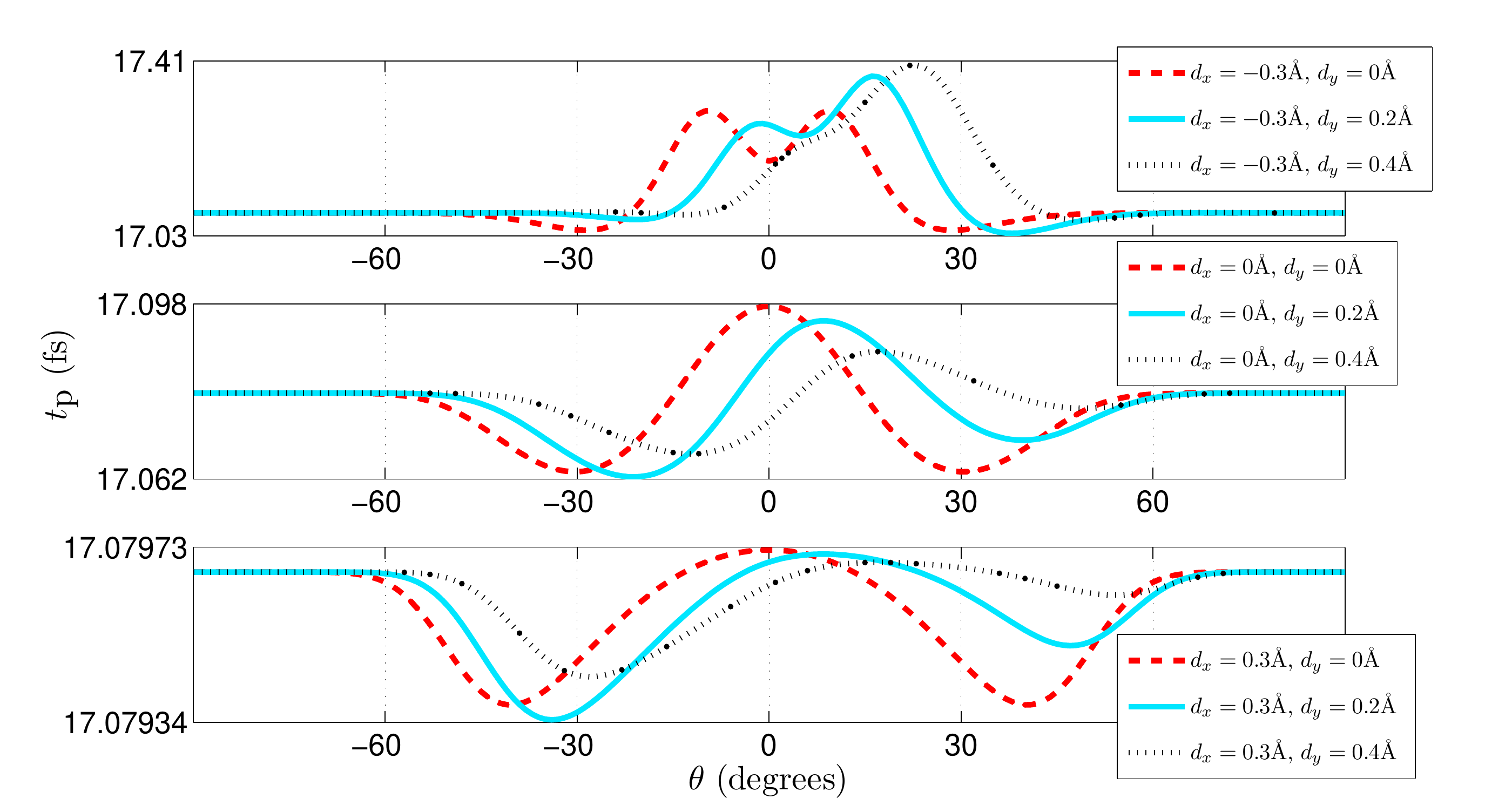}
	\caption{$\gamma = 0.55$.}
	\label{figure04a}
	\end{subfigure}
	\begin{subfigure}{.85\textwidth}
	\centering
	\includegraphics[width=\linewidth]{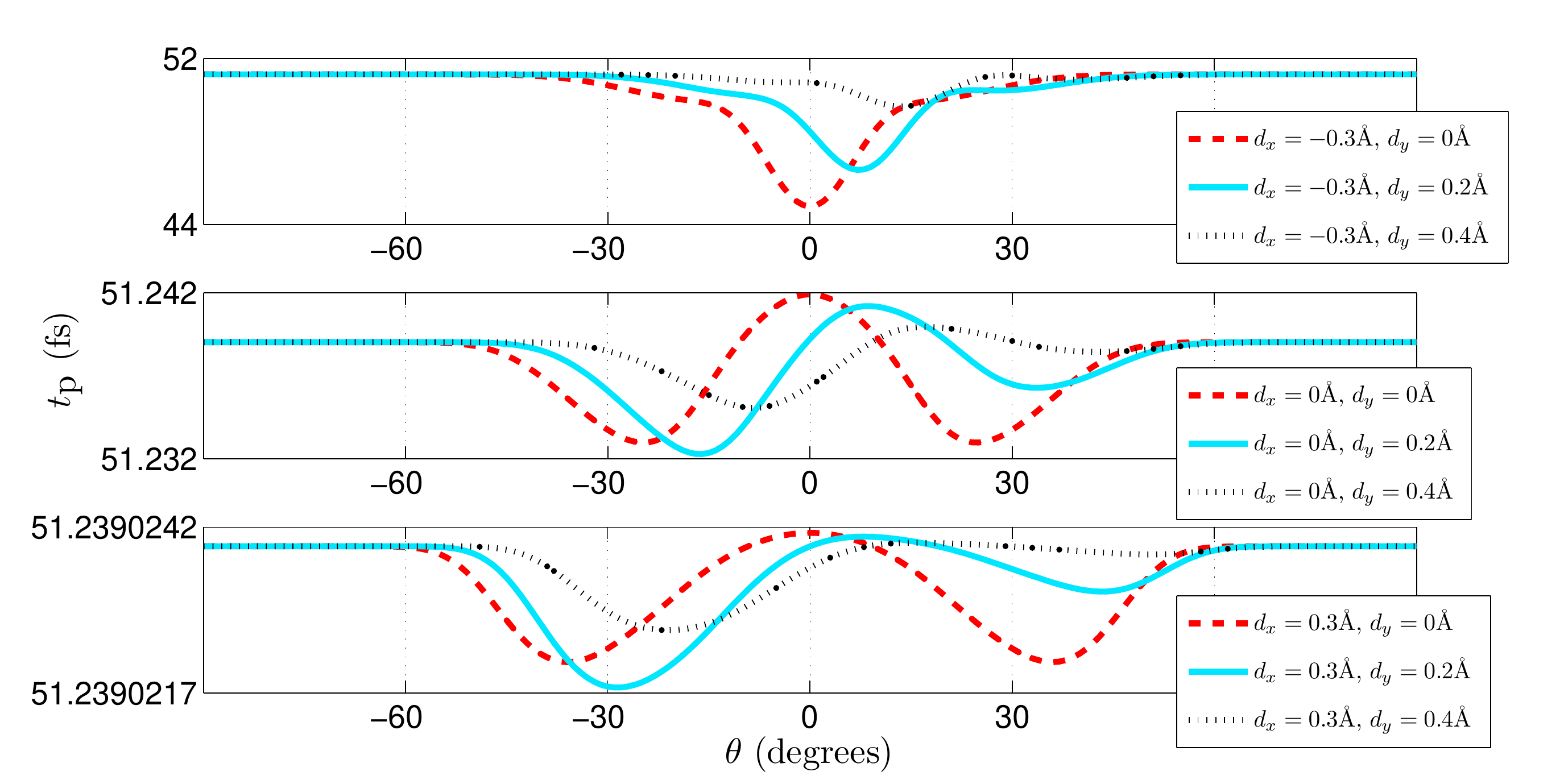}
	\caption{$\gamma = 0.85$.}
	\label{figure04b}
	\end{subfigure}
	\captionsetup{width=0.75\textwidth}
	\caption{$t_\textnormal{p}$ as functions of $\theta$, given various combinations of $(\gamma, d_x, d_y)$.}
	\label{figure04}
	\end{figure}

Furthermore, our results show that for every $(\gamma, d_x)$, we have 
	\begin{align}	
	t_\textnormal{p} (\gamma, d_x, d_y, \theta) = t_\textnormal{p} (\gamma, d_x, -d_y, -\theta). \label{tp_sym}
	\end{align}
This is because a deformation consisting of a shift of $d_y$ and rotation of $\theta$ is intrinsically identical to one consisting of a shift and rotation of the same magnitudes but both in the opposite direction. \Cref{figure04} shows variations in $t_\textnormal{p}$ as $\theta$ varies between $-90^\circ$ and $90^\circ$, whilst $(\gamma, d_x, d_y)$ are fixed at certain values. For every combination of $(\gamma, d_x)$, we have presented only results relating to $d_y \ge 0$, since one can simply reflect these curves about $\theta = 0$ to obtain results for $d_y < 0$. For fixed $(\gamma, d_x)$ with $d_y = 0$, the graph of $t_\textnormal{p}(\theta)$ is symmetric about $\theta = 0$, where the graph has a local mimimum under some $(\gamma, d_x)$ and a local maximum under others; we find from our results that for every $\gamma$ there is one value of $d_x$ at which the graph transitions from having a local minimum to having a local maximum at $\theta = 0$, and that this value of $d_x$ increases with $\gamma$. For fixed $(\gamma, d_x)$ with $d_y \neq 0$, the symmetry of $t_\textnormal{p}(\theta)$ about $\theta = 0$ is broken, and as $d_y$ increases, the local extremum which was at $\theta = 0$ when $d_y = 0$ moves towards larger $\theta$. There are cases where this local extremum ceases to exist when $d_y$ becomes large, for instance the case of $(\gamma, d_x) = (0.55, -0.3\textnormal{\r{A}})$, as we can see in \Cref{figure04a}: there is a local minimum at $\theta = 0$ if $d_y = 0$ and at $\theta = 5^\circ$ if $d_y = 0.2 \textnormal{\r{A}}$, but if $d_y = 0.4 \textnormal{\r{A}}$ then this local mimimum disappears. For any fixed $(\gamma, d_x, d_y)$, we always have $t_\textnormal{p}$ tending to some value as $\theta$ tends to $\pm 90^\circ$, typically with several local extrema between $\theta = 0$ and $\theta = \pm 90^\circ$; the value of this limit at $\pm 90^\circ$ is dependent only on $\gamma$. Calling this limit  $t_\textnormal{p}^{90}(\gamma)$, we have  $t_\textnormal{p}^{90} (0.55) = 17.1 $fs, and $t_\textnormal{p}^{90} (0.85) = 51.2$fs. As $\gamma \rightarrow 1$ and as $\gamma \rightarrow 0$, we have $t_\textnormal{p}^{90} (\gamma) \sim t_\textnormal{p}^\textnormal{max} (\gamma)$, and for $0.02 \lessapprox \gamma \lessapprox 0.4$, we have $t_\textnormal{p}^{90} (\gamma) \sim t_\textnormal{p}^\textnormal{min} (\gamma)$.
	\begin{figure}[h!]
	\centering
	\begin{subfigure}{.33\textwidth}
	\centering
	\includegraphics[width=\linewidth]{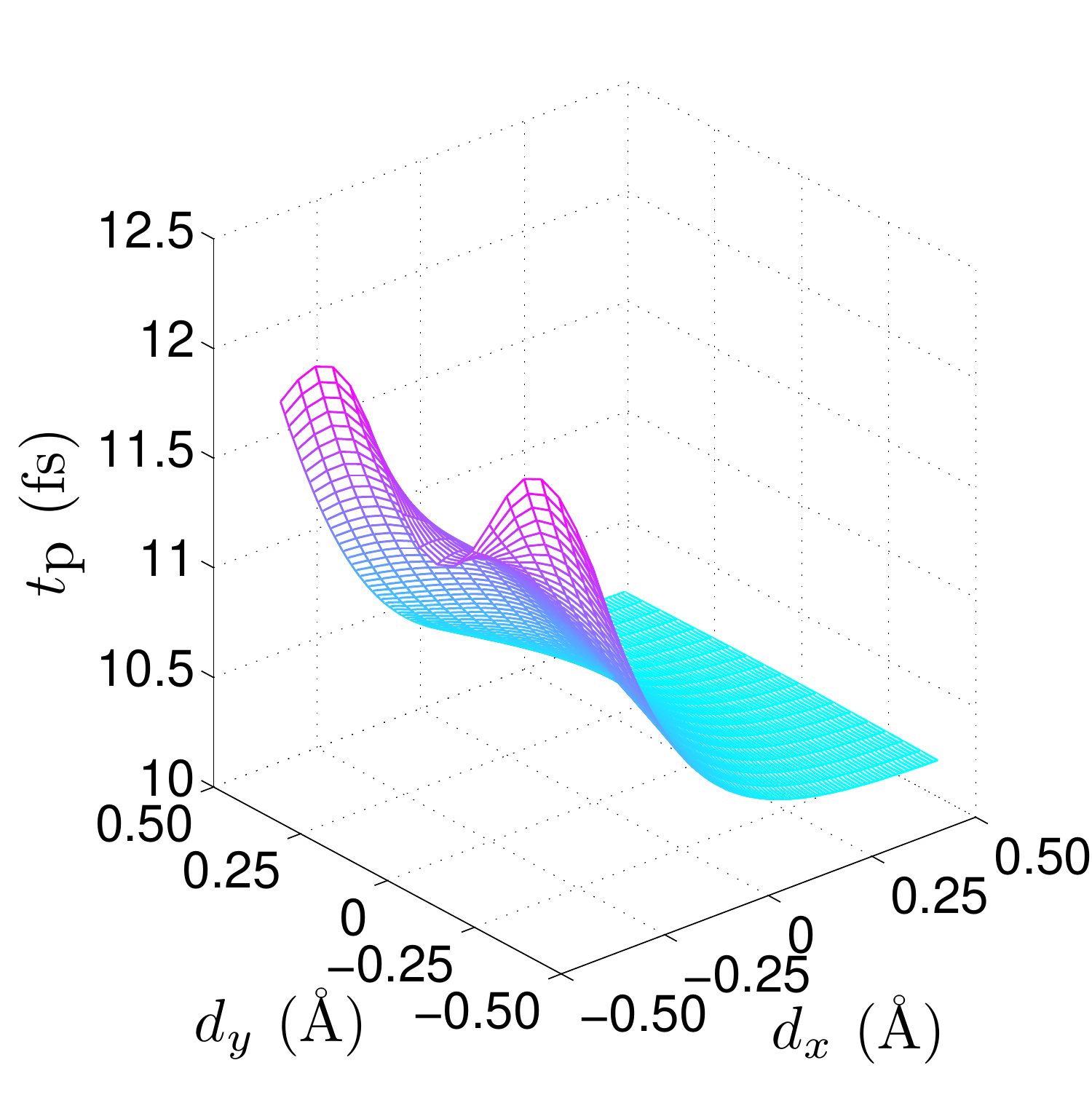}
	\caption{$\gamma = 0.25, \theta = 0$.}
	\label{figure05a}
	\end{subfigure}%
	\begin{subfigure}{.33\textwidth}
	\centering
	\includegraphics[width=\linewidth]{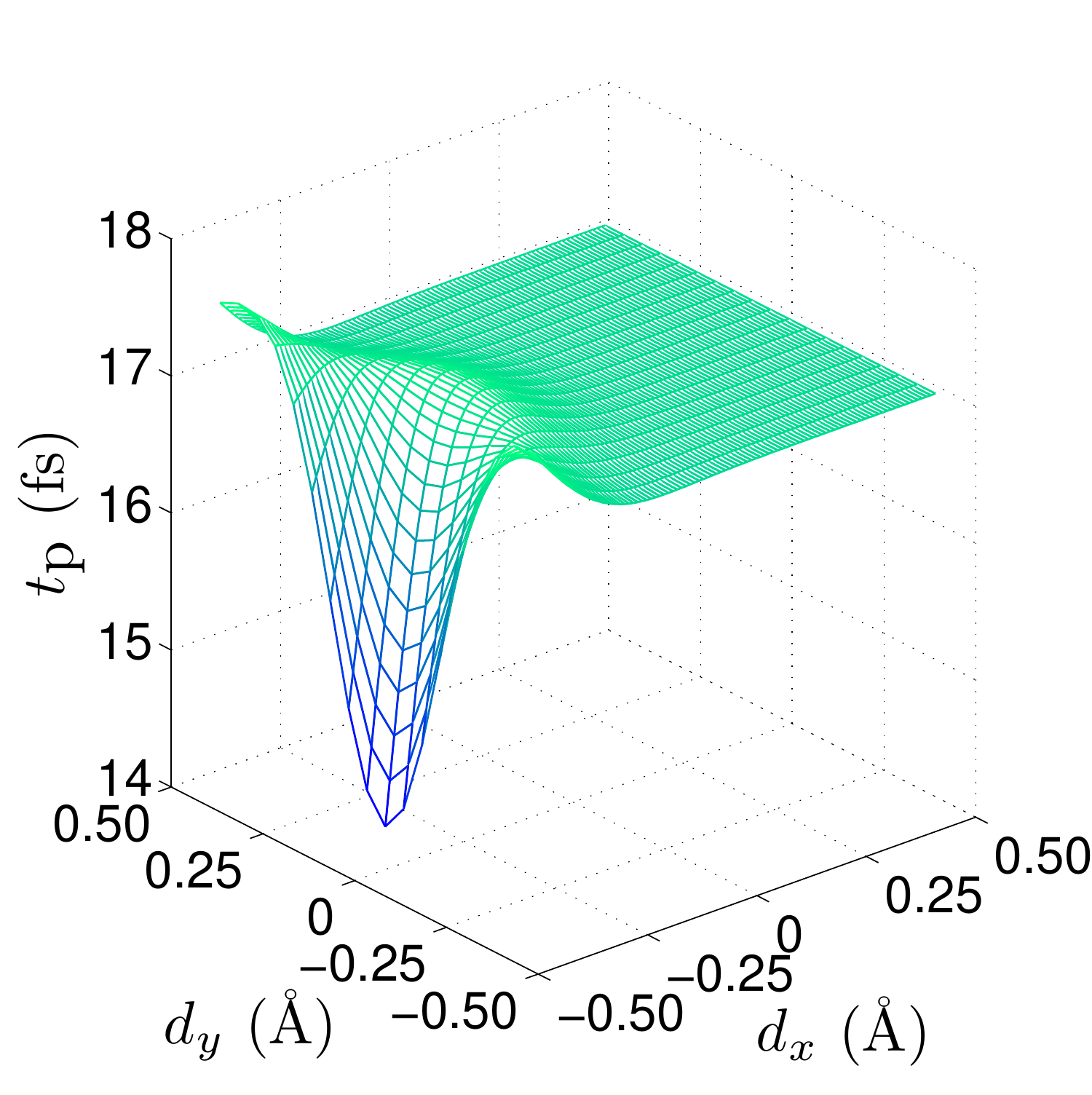}
	\caption{$\gamma = 0.55, \theta = 0$.}
	\label{figure05b}
	\end{subfigure}%
	\begin{subfigure}{.33\textwidth}
	\centering
	\includegraphics[width=\linewidth]{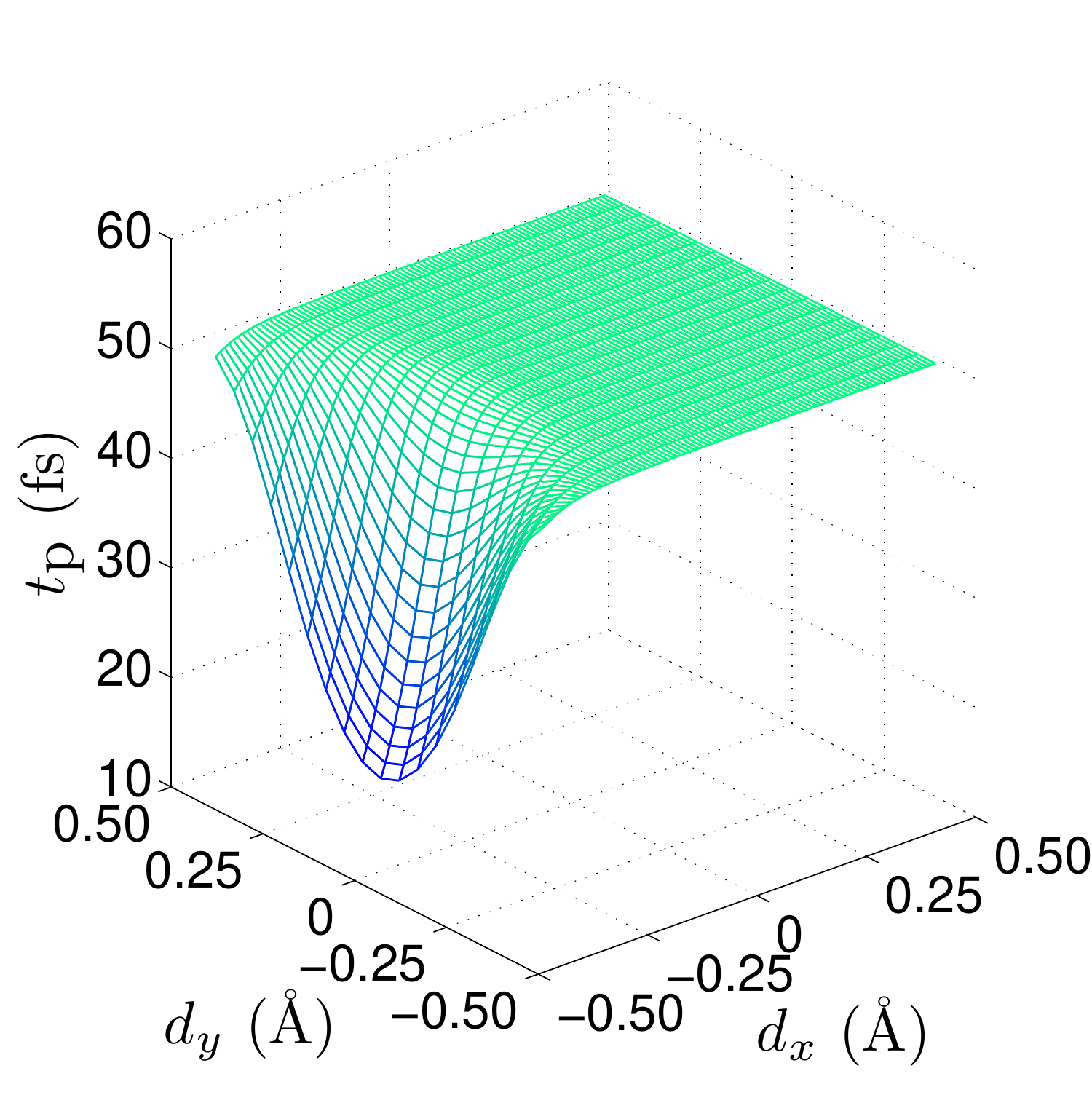}
	\caption{$\gamma = 0.85, \theta = 0$.}
	\label{figure05c}
	\end{subfigure}
	\centering
	\begin{subfigure}{.33\textwidth}
	\centering
	\includegraphics[width=\linewidth]{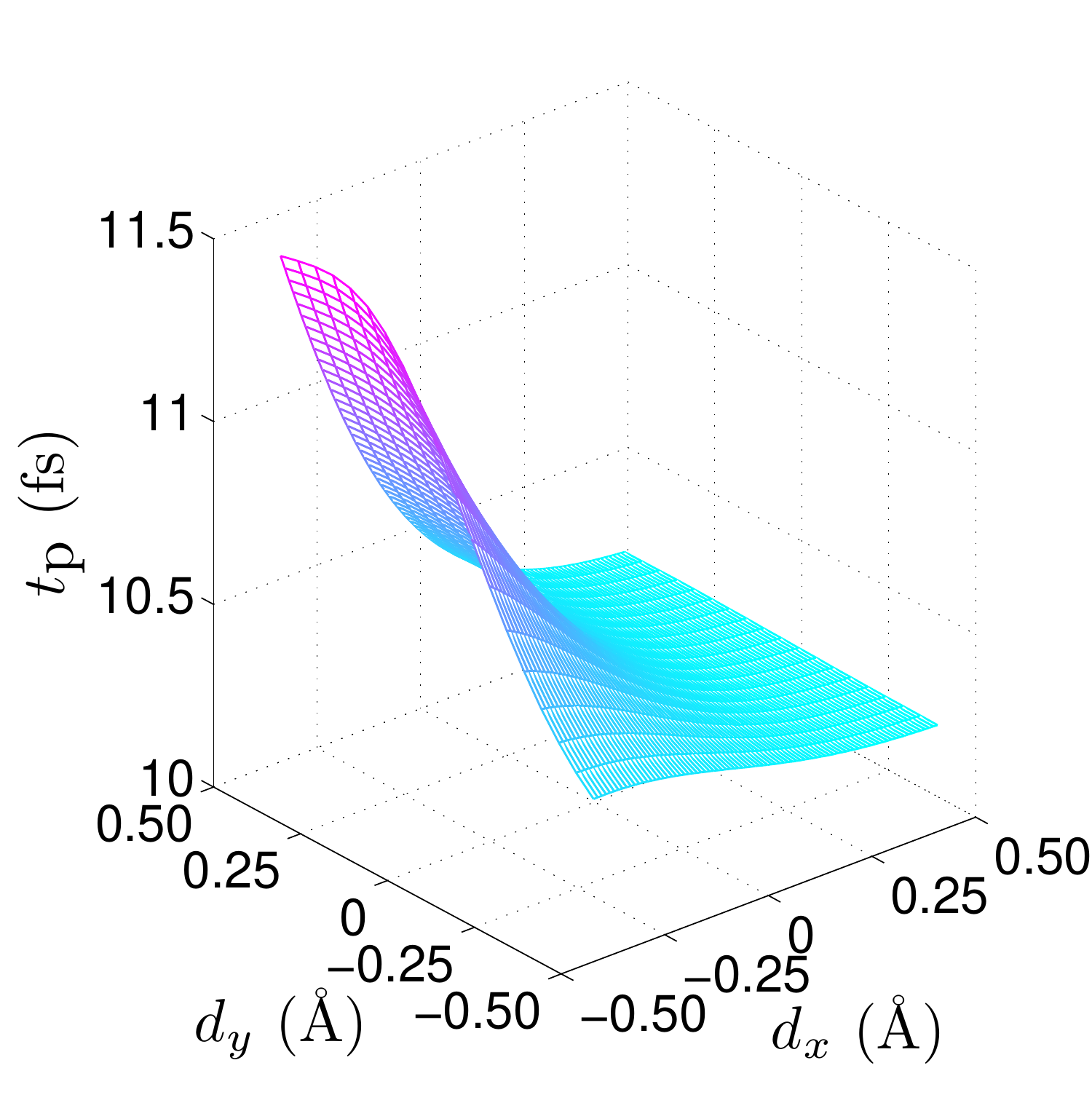}
	\caption{$\gamma = 0.25, \theta = 40^\circ$.}
	\label{figure05d}
	\end{subfigure}%
	\begin{subfigure}{.33\textwidth}
	\centering
	\includegraphics[width=\linewidth]{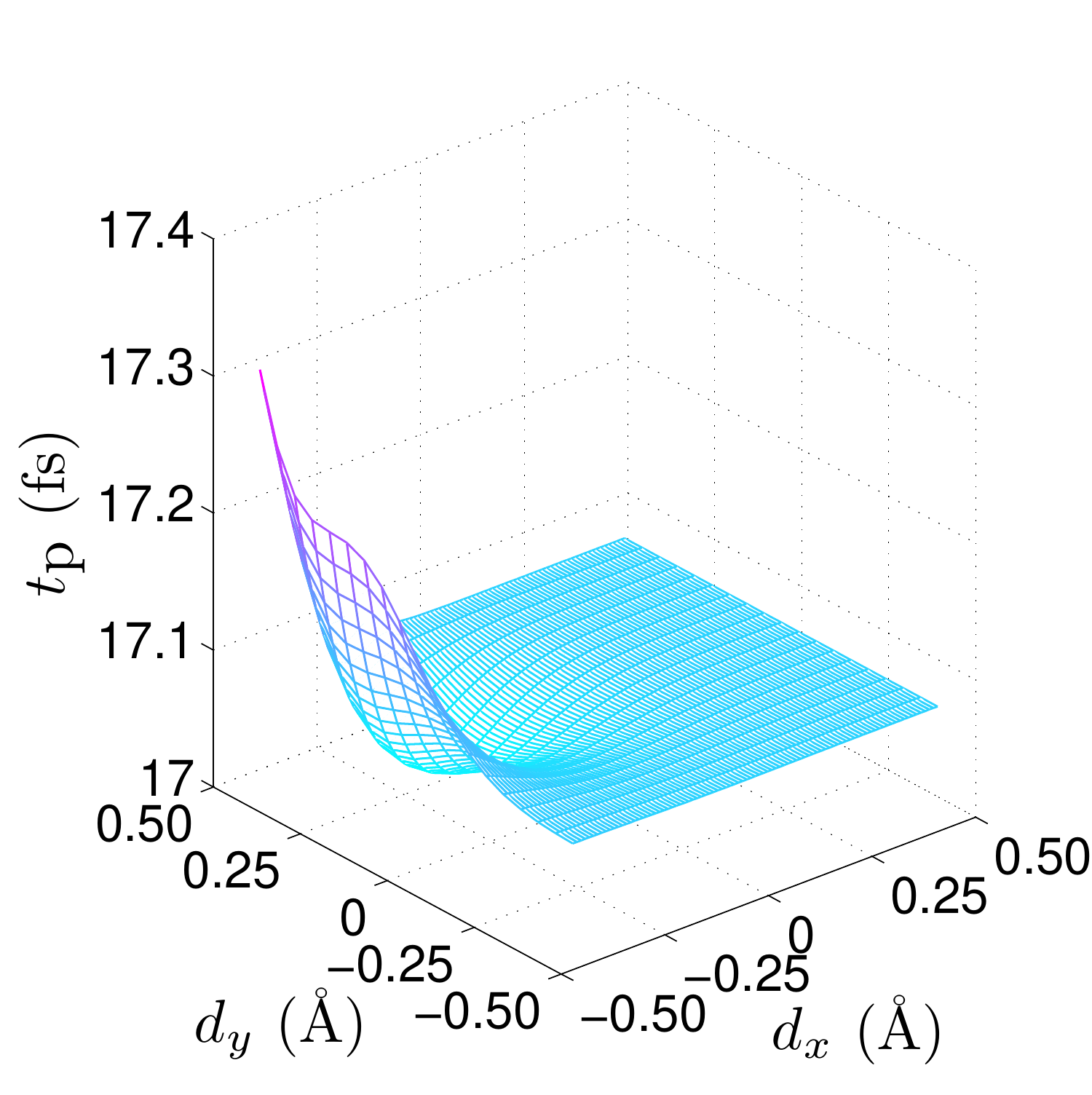}
	\caption{$\gamma = 0.55, \theta = 40^\circ$.}
	\label{figure05e}
	\end{subfigure}%
	\begin{subfigure}{.33\textwidth}
	\centering
	\includegraphics[width=\linewidth]{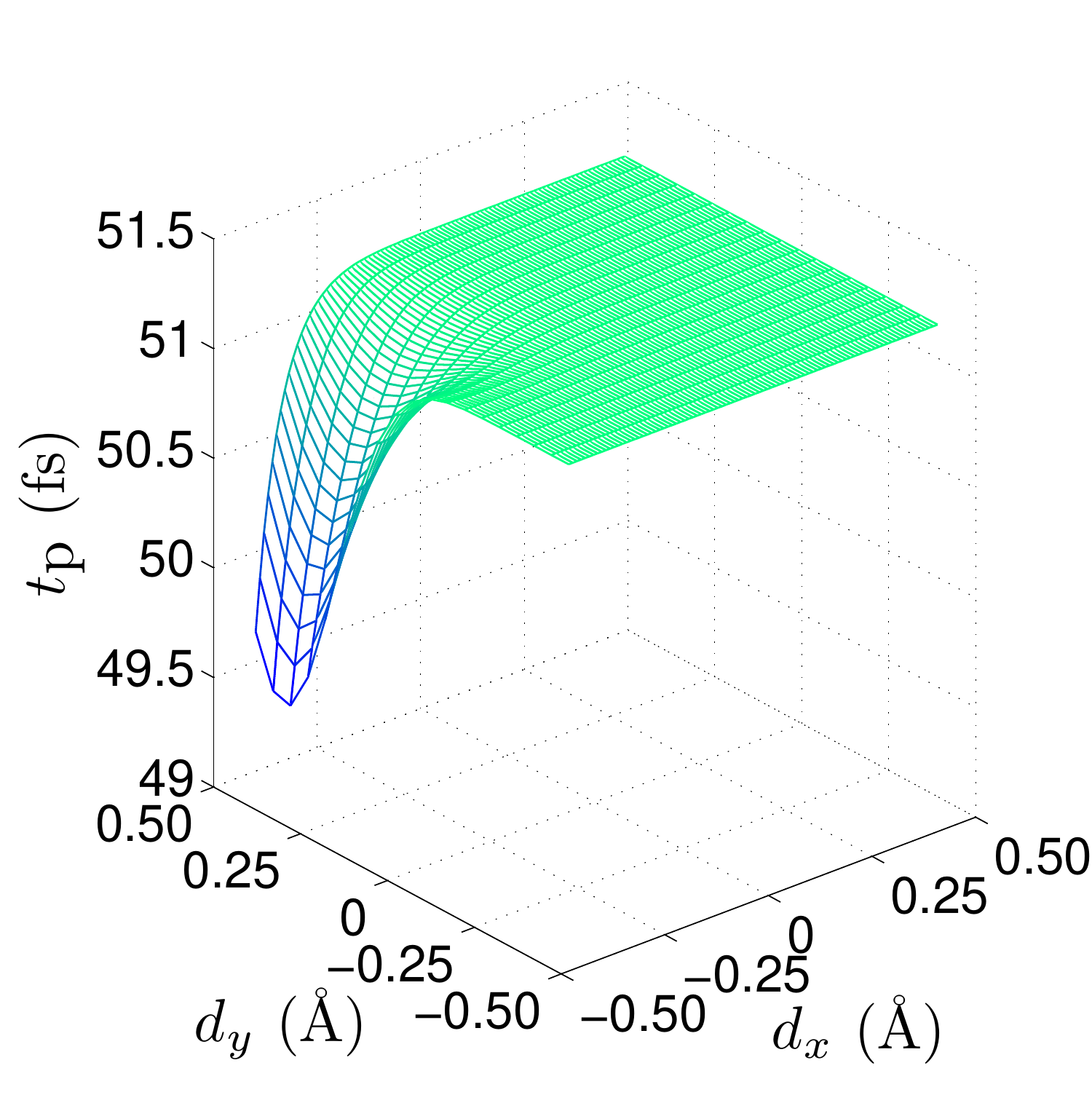}
	\caption{$\gamma = 0.85, \theta = 40^\circ$.}
	\label{figure05f}
	\end{subfigure}
	\captionsetup{width=0.75\textwidth}
	\caption{$t_\textnormal{p}$ as surfaces over the parameter subspace $(d_x, d_y)$, given various combinations of $(\gamma, \theta)$. In each case, the range of $d_x$ is $d_x^\textnormal{crit} \le d_x \le 0.45\textnormal{\r{A}}$.}
	\label{figure05}
	\end{figure}

We further observe by comparing \Cref{figure04a,figure04b} that, when $\gamma = 0.85$, there is a larger overall variation in $t_\textnormal{p}$ as a result of varying $(d_x,d_y,\theta)$, compared to when $\gamma = 0.55$. This agrees with our observation about \Cref{figure03} that the gap between $t_\textnormal{p}^\textnormal{min} (\gamma)$ and $t_\textnormal{p}^\textnormal{max} (\gamma)$ increases as $\gamma \rightarrow 1$. Indeed, this gap also increases as $\gamma \rightarrow 0$. Moreover, for fixed $\gamma$, the larger $d_x$ is, the less $t_\textnormal{p}$ varies with $\theta$ or with $d_y$. As we see in \Cref{figure05b,figure05c,figure05e,figure05f}, if $\gamma$ is far from 0, then for fixed $(\gamma, \theta)$, $t_\textnormal{p}$ as a surface over $(d_x, d_y)$ is almost constant given sufficiently large $d_x$. As $d_x \rightarrow \infty$, $t_\textnormal{p}$ always tends to some limit, whose value is independent of $d_y$. Meanwhile, we see in \Cref{figure05a,figure05b,figure05c} that if $\theta = 0$, then for fixed $(\gamma, \theta)$, $t_\textnormal{p}$ as a surface over $(d_x, d_y)$ is symmetric about the line $d_y = 0$. This is due to \cref{tp_sym}. If $\theta = 0$ and $\gamma$ is moderate, such as 0.55, then for each $d_y$ sufficiently to 0 we have some small value of $d_x$ which maximises $t_\textnormal{p}$, as we can see in \Cref{figure05b}. This shows that increasing $d_x$, which represents moving the donor away from the acceptor in the H bond, does not necessarily prolong the proton tunnelling. If $\theta \neq 0$, then the symmetry about $d_y = 0$ is broken, and reflecting a surface for $\theta > 0$ about the line $d_y = 0$ produces corresponding results for $\theta < 0$.

	\section{Discussions and Conclusions} \label{conclusions}

We have studied the quantum mechanical tunnelling of a proton across the potential barrier between the donor and acceptor of a planar hydrogen bond in DNA, and computed an analytical expression for the proton's characteristic tunnelling time (CTT) as a function of four parameters describing the geometry of the bond. Three of these parameters, $d_x, d_y$ and $\theta$, represent the deformation of the H bond from its normal alignment, under the assumption that any deformation consists of planar translations and rotations of the donor and acceptor molecules as independent units. With the acceptor molecule treated without loss of generality as fixed,  $d_x$ and $d_y$ respectively represent the longitudinal and lateral displacements of the donor molecule from its normal position, while $\theta$ represents the rotation of the donor molecule about the donor atom from its normal orientation. The fourth parameter, $\gamma$, taking values $0 < \gamma \le 1$, represents the intrinsic symmetry that the potential in the H bond possesses when the bond is in its normal alignment. When $\gamma = 1$, we recover a model previously studied in \cite{Krasilnikov2014}, whose potential function in the normal H bond was symmetric about the potential barrier, so that the local potential wells near the donor and acceptor are equivalent to each other. This symmetry is broken only if some of $(d_x, d_y, \theta)$ is non-zero. For $0 < \gamma < 1$, the symmetry is broken even if $d_x = d_y = \theta = 0$, in the sense that the local potential well near the donor has a less energetic ground state than the one near the acceptor, and this gives a better representation of the physical property of the H bond than $\gamma = 1$. In addition, setting any of $d_x, d_y$ and $\theta$ to non-zero values further distorts the symmetry between the two local potential wells. 

We have discovered that some combinations of $(\gamma, d_x, d_y, \theta)$ provide potential functions which cannot model a tunnelling process, because the potential barrier is not higher than the ground state energy of a proton in equilibrium near the donor. The smaller $\gamma$ is, the more $(d_x, d_y, \theta)$ combinations provide invalid models, meaning that the region of validity in our parameter space shrinks as $\gamma$ decreases. For $0.01 \le \gamma \le 0.99, -0.45\textnormal{\r{A}} \le d_x, d_y \le 0.45 \textnormal{\r{A}}, -90^\circ \le \theta \le 90^\circ$, and excluding all invalid parameter combinations, we have found that $8.5\textnormal{fs} \le t_\textnormal{p} (\gamma, d_x, d_y, \theta) \le 770$fs, where $t_\textnormal{p}$ stands for the proton's CTT. For each $\gamma$, certain $(d_x, d_y, \theta)$ combinations minimise or maximise $t_\textnormal{p}$, and we have found that  $t_\textnormal{p}^\textnormal{min} (\gamma) $ is a slowly-varying function taking values around 10fs, whilst  $t_\textnormal{p}^\textnormal{max} (\gamma)$ diverges as $\gamma \rightarrow 0$ and grows rapidly towards $\mathcal{O}(10^{27})$s as $\gamma \rightarrow 1$. Taking the mean $t_p$ over all $(d_x, d_y, \theta)$ for every fixed $\gamma$, we have found that  $t_\textnormal{p}^\textnormal{ave} (\gamma) \sim t_\textnormal{p}^\textnormal{max} (\gamma)$ as $\gamma \rightarrow 1$. This means that in an H bond selected at random from a statistical ensemble, the proton's CTT is likely to be as large as it can be if the potential in the bond has a high $\gamma$-symmetry. On the other hand, we have also observed that if $\gamma$ takes moderate values such as $\gamma \approx 0.3$, then $t_\textnormal{p}^\textnormal{ave} (\gamma) \sim t_\textnormal{p}^\textnormal{min} (\gamma)$, meaning that the proton's CTT is likely to be as small as it can be in this case. As $\gamma \rightarrow 0$, $t_\textnormal{p}^\textnormal{ave} (\gamma)$ is not asymptotic to $t_\textnormal{p}^\textnormal{min} (\gamma)$ or $t_\textnormal{p}^\textnormal{max} (\gamma)$; given the fact that $t_\textnormal{p}^\textnormal{max} (\gamma)$ diverges towards infinity in this case, we deduce that parameter combinations resulting in large $t_\textnormal{p}$ are rare when $\gamma$ is small. We have investigated how $t_\textnormal{p}$ varies with $\theta$ given fixed $(\gamma,d_x,d_y)$, and found that as $\theta \rightarrow \pm 90^\circ$, $t_\textnormal{p}$ always converges to some $t_\textnormal{p}^{90} (\gamma)$ which depends on $\gamma$ in the following manner.  In extreme cases of $\gamma \rightarrow 1$ and $\gamma \rightarrow 0$, we have $t_\textnormal{p}^{90} (\gamma) \sim t_\textnormal{p}^\textnormal{max} (\gamma)$, and for moderate $\gamma$ values, we have $t_\textnormal{p}^{90} (\gamma) \sim t_\textnormal{p}^\textnormal{min} (\gamma)$. For $-90^\circ < \theta < 90^\circ$, we have observed that $t_\textnormal{p}$ has various local maxima and local minima but the variation in $t_\textnormal{p}$ is small unless either $\gamma$ is close to extremal values, or $d_x$ is negative with large magnitudes. For example, if $0.3 \le \gamma \le 0.99$ and $d_x \ge 0$, then regardless of $d_y$, we have the result that as $\theta$ varies, $t_\textnormal{p}$ never deviates by more than 1\% from some average value. We have also investigated how $t_\textnormal{p}$ varies with $(d_x, d_y)$, given fixed $(\gamma , \theta )$, and found that if $d_x$ is sufficiently large, then $t_\textnormal{p}$ is an almost-constant surface over $(d_x, d_y)$, and that $t_\textnormal{p}$ tends to some $d_y$-independent limit as $d_x \rightarrow \infty$. Since large $d_x$ corresponds to large donor-acceptor separation, one might expect $t_\textnormal{p}$ to be maximised in the limit $d_x \rightarrow \infty$, but our results show that this is not always the case. 

The most important difference that generalising from $\gamma = 1$ to $0 < \gamma \le 1$ has made is that, for most $\gamma$ values in $0 < \gamma < 1$, the proton CTT is sub-picosecond regardless of $(d_x, d_y, \theta)$. Compared to the $\gamma = 1$ case in which some $(d_x, d_y, \theta)$ give CTTs of $\mathcal{O}(10^{27})$s, the sub-picosecond time-scale is much more biologically relevant. Moreover, if $\gamma$ is such that the CTT is guaranteed to be sub-picosecond, then it varies by no more than 2 orders of magnitude as the H bond deforms. This means that the tunnelling process is much more stable with respect to bond deformation compared to the $\gamma = 1$ case, under which the CTT varies by over 30 orders of magnitude as the H bond deforms. Overall, our model under moderate $\gamma$-values produces CTTs on a biological time-scale with strong stability against bond deformation, and therefore it supports the theory that proton tunnelling across DNA hydrogen bonds may be a mechanism responsible for biological processes such as spontaneous mutation. 

	\vspace{\baselineskip}
The author is grateful to Dr. Emma Coutts and Dr. Bernard Piette for their kind support.
	
	\vspace{\baselineskip}
	\subsection*{Appendix}

In \Cref{model} we presented the overlap integral $S$ and transition integrals $I_{jk}$, for $j,k = 1,2$ [cf. \cref{defn_integrals,Iij_details}]. We have computed closed-form expressions for these integrals, as follows.
	\begin{subequations}
	\begin{align}
	S &= 2 ~\frac{\sqrt{ g \gamma }}{K_{0,12}} \exp \left[ K_{1,12} + \frac{K_{2,12}^2}{2 B_{12} K_{0,12}^2} \right] , \\
	I_{jk} &= \sqrt{g} \gamma^{\frac{ j + k }{2} - 1} \left[ \left( \frac{b Q_{jk}^2}{B_{jk}^2} + \frac{b - 2 q Q_{jk}}{B_{jk}} \right) J_{0,jk} \right. \nonumber \\
	& \qquad  \qquad \qquad + 2 (-1)^{ k - 1 } \left( \frac{b C_{jk} Q_{jk}}{B_{jk}^2} - \frac{\left( c Q_{jk} + q C_{jk} \right)}{B_{jk}} + p \right) J_{1,jk} \nonumber \\
	& \qquad  \qquad \qquad \left. + \left( \frac{b C_{jk}^2}{B_{jk}^2} - \frac{2 c C_{jk}}{B_{jk}} + a \right) J_{2,jk} \right],
	\end{align}
	\end{subequations}
where
	\begin{subequations}
	\begin{align}
	J_{0,jk} &= \frac{1}{2 K_{0,jk}} \exp \left( K_{1,jk} + \frac{K_{2,jk}^2}{2 B_{jk} K_{0,jk}^2} \right) \textnormal{erfc} \left( \frac{K_{2,jk}}{(-1)^{ k - 1 } \sqrt{2 B_{jk}} K_{0,jk}} \right), \\
	J_{1,jk} &= \frac{\sqrt{B_{jk}}}{\sqrt{2 \pi} K_{0,jk}^2} \exp \left( K_{1,jk} \right) \nonumber \\
	& \quad + (-1)^k \frac{K_{2,jk}}{2 K_{0,jk}^3} \exp \left( K_{1,jk} + \frac{K_{2,jk}^2}{2 B_{jk} K_{0,jk}^2} \right) \textnormal{erfc} \left( \frac{K_{2,jk}}{(-1)^{ k - 1 } \sqrt{2 B_{jk}} K_{0,jk}} \right), \\
	J_{2,jk} &= (-1)^k \frac{\sqrt{B_{jk}} K_{2,jk}}{\sqrt{2 \pi} K_{0,jk}^4} \exp \left( K_{1,jk} \right) \nonumber \\
	& \quad + \frac{\left( K_{2,jk}^2 + B_{jk} K_{0,jk}^2 \right)}{2 K_{0,jk}^5} \exp \left( K_{1,jk} + \frac{K_{2,jk}^2}{2 B_{jk} K_{0,jk}^2} \right) \textnormal{erfc} \left( \frac{K_{2,jk}}{(-1)^{ k - 1 } \sqrt{2 B_{jk}} K_{0,jk}} \right) ,
	\end{align}
	\end{subequations}
with
	\begin{align}
	K_{0,jk} = \sqrt{A_{jk} B_{jk} - C_{jk}^2}, \quad K_{1,jk} = \frac{Q_{jk}^2}{2B_{jk}} - \frac{R_{jk}}{2}, \quad K_{2,jk} = B_{jk} P_{jk} - C_{jk} Q_{jk},
	\end{align}
and $\textnormal{erfc}$ being the cumulative error function, defined for all real $X$ by 
	\begin{align}
	\textnormal{erfc}(X) = (2 / \sqrt{\pi}) \int_X^\infty e^{-z^2} \textnormal{d} z.
	\end{align}
The parameters $a,b,c,p,q,A_{jk},B_{jk},C_{jk},P_{jk},Q_{jk},R_{jk}$ were defined in the main text.



\end{document}